\documentclass[pra,showpacs,twocolumn,superscriptaddress]{revtex4}

\usepackage{graphicx}
\usepackage{epsfig}
\usepackage{dcolumn}
\usepackage{bm}
\usepackage{color}

\newcommand{\bra}[1]{\left\langle #1 \right|}
\newcommand{\ket}[1]{\left| #1 \right\rangle}

\newcommand{\ketbra}[2]{\left|#1\middle\rangle\middle\langle#2\right|}

\newcommand{\M}[1]{\mathcal{#1}}
\newcommand{\Md}[2]{\mathcal{#1}^{(#2)}}

\begin{document}

\title{Quantifying Quantum Correlations in Fermionic Systems using Witness Operators}
\author{Fernando Iemini}
\email{iemini81@yahoo.com.br}
\author{Thiago O. Maciel}
\email{thiago@fisica.ufmg.br}
\author{Tiago Debarba}
\email{debarba@fisica.ufmg.br}
\author{Reinaldo O. Vianna}
\email{reinaldo@fisica.ufmg.br}
\affiliation{Departamento de F\'{\i}sica - ICEx - Universidade Federal de Minas Gerais,
Av. Pres.  Ant\^onio Carlos 6627 - Belo Horizonte - MG - Brazil - 31270-901.}

\date{\today}

\begin{abstract} 
We present a method to quantify  quantum correlations 
in arbitrary systems of indistinguishable fermions using witness operators. 
The method associates the problem of finding the optimal entanglement 
witness  of a  state with a class of problems known as semidefinite programs
(SDPs), which can be solved efficiently with arbitrary accuracy. 
Based on these optimal witnesses, we introduce a measure of quantum 
correlations which has an interpretation analogous to the
 Generalized Robustness of entanglement. 
We also extend the notion of quantum discord 
to the case  of indistinguishable fermions, 
and propose a geometric quantifier, which is compared to our entanglement
measure.
Our numerical results show a remarkable equivalence between 
the proposed Generalized Robustness and the Schliemann concurrence, 
which are equal for 
pure states. For mixed states, the Schliemann concurrence presents 
itself as an upper bound for the Generalized Robustness.
The quantum discord is also found to be an upper bound for the 
entanglement.
\end{abstract}

\maketitle




\section{Introduction}

The notion of entanglement, first noted by Einstein, Podolsky and 
Rosen \cite{epr35}, is considered one of the main features of quantum 
mechanics, being subject of study in several areas recently
\cite{Horodecki09, Vedral08, Kais}. 
Thus, it  is of paramount importance both the  understanding and
 the quantification of entanglement in composite quantum systems,
 being these one of the main 
challenges of modern quantum theory. 

Despite being a subject widely studied in systems of non-identical particles,
or whose particles, though identical,  are 
well separated from each other, being thus distinguishable, 
less attention was given to the case where this separation is 
very small, such that the overlap of  their wave functions is no longer 
negligible. 
In this case we have to take into account the indistinguishability of 
the particles, being the space of quantum states restricted to symmetric or 
antisymmetric subspaces of the Hilbert space, depending on the bosonic or 
fermionic nature of the system.
The study of entanglement for systems of indistinguishable particles has 
been, however, a subject of great controversy, leading to different 
approaches in its treatment. Among the most mentioned we have: quantum 
correlations \cite{Schliemann02}, entanglement of modes \cite{zanardi02},
 entanglement of particles \cite{wiseman03}. 
The concept of quantum correlations is based on the notion that the 
correlations generated by mere (anti-)symmetrization of the state 
due to indistinguishability of their particles do not constitute 
truly as entanglement. 
We will analyse, in this paper, the quantum correlations in systems 
of indistinguishable fermions.

The entanglement in  two-fermion sates 
  with a four-dimensional single-particle space 
($\M{A}(\Md{H}{4}\otimes\Md{H}{4})$) can be characterized by the
 Schliemann concurrence \cite{Schliemann02}.
For pure states in arbitrary dimension, one can use
 the  von Neumann entropy 
of the reduced single-particle density matrix as a measure
of entanglement \cite{paskauskas01}. 
However, the problem of quantification or even detection of entanglement
in the  general 
case is still open  \cite{ghirardi04,plastino09,plastino10}.
A useful  concept  is that of
 entanglement witness \cite{witness}:
 a Hermitian
operator with  non-negative expectation value for all 
separable states, but which can have a negative expectation
value for an entangled state.
We will focus on 
 optimal entanglement witnesses 
(OEW), that can be used not only to {\em witness}  the entanglement,
 but also 
to quantify it \cite{brandao05,Reinaldo04b,rov06}.
In this paper, we will see  how to determine such 
OEWs, and especially how to use them to quantify the quantum correlations 
in fermionic systems. We will also confront our measure of entanglement
with the {\em quantum discord}  \cite{oz2001,hv2001,bm2011,dvb2010}.
In order to do that, we will define a quantum discord for fermionic
particles. Previous studies of quantum discord in systems of indistinguishable
particles, like in \cite{wang2010}, employ the notion of correlation of
 modes \cite{zanardi02}, which are distinguishable.

This paper is organized as follows. 
In Sec.\ref{backgrounds} we recall some known concepts and tools of 
the theory of entanglement, as the witnessed entanglement and the 
quantum correlations, which will be essential to the development of 
the ideas throughout the article. 
In Sub-sec.\ref{witnessed.entanglement}, we recall the definition of 
optimal entanglement witness, and briefly discuss  its use as 
an entanglement quantifier. 
In Sub-sec.\ref{quantum.correlations} we recall the ideas associated with 
the concept of quantum correlations, as well as the  
Schliemann concurrence.
In Sub-sec.\ref{discord} we revise the concept of
quantum discord for distinguishable particles.  
In Sec.\ref{quantifying.quantum.correl} we introduce our method for quantifying 
the quantum correlations in fermionic systems using witness operators, 
and define the Fermionic Generalized Robustness. We also extend the notion
of quantum discord to fermions, such that it takes into account the
particles' indistinguishability,  and introduce the 
{\em Fermionic Geometric Discord}.
In Sec.\ref{numerical.results} we show  numerical results, comparing
the Fermionic Generalized Robustness, the Schliemann concurrence,
and the Fermionic Geometric Discord.
We finish the illustration of the method with a beautiful 
{\em quantum phase diagram} yielded by the five-partite  Fermionic
Generalized Robustness in the Extended Hubbard Model.
We conclude in Sec.V.

\section{Preliminary Concepts}
\label{backgrounds}

We will see, in this section, some familiar concepts  from  the theory of 
entanglement,  namely, {\em witnessed entanglement} 
\cite{brandao05, Reinaldo04b, rov06}, 
quantum correlations \cite{Schliemann02} 
and quantum discord \cite{oz2001,hv2001,bm2011,dvb2010}.
 A reader already familiarized with such concepts might skip to the next 
section of the article.

\subsection{Witnessed Entanglement}
\label{witnessed.entanglement}

Entanglement witnesses  are Hermitian operators (observables - $W$) whose 
expectation values contain information about the entanglement of 
quantum states. 
The operator $W$ is an entanglement witness for a given entangled 
quantum state $\rho$ if the following conditions are satisfied 
\cite{witness}: its 
expectation  value is negative for the particular entangled quantum state 
($Tr(W\rho) < 0 $), while it is non-negative  on the set of separable 
states ($S$) 
($\forall \sigma \in \M{S}, \,\,\, Tr(W\sigma)\geq 0 $).
We are particularly  interested in  optimal entanglement witnesses. 
\textit{$W_{opt}$} is the OEW
for the state $\rho$ if 
\begin{equation}
Tr(W_{opt}\rho) = \min\limits_{W \in \M{M}} \,Tr(W\rho),
\label{optimal.witness.definition}
\end{equation}
where $\M{M}$ represents a compact subset of the set of 
entanglement witnesses $\M{W}$.

OEWs can be used to quantify  entanglement.
Such quantification is related to  the choice of the set $\M{M}$, where 
different sets will determine different quantifiers \cite{brandao05}. 
We can define these quantifiers by:
\begin{equation}
E(\rho) = max(0 , -\min\limits_{W \in \M{M}} \,Tr(W\rho)).
\label{OEW}
\end{equation}

An example of a quantifier that can be calculated using OEW is the 
 Generalized Robustness of entanglement \cite{vidal99} ( $\M{R}_g(\rho)$), 
which is defined as the minimum required mixture such that 
a separable state is obtained.
Precisely, it is the minimum value of $s$ such that
\begin{equation}
\sigma=\frac{\rho + s\varphi}{1+s}
\label{rob.generalized.definition}
\end{equation} 
be a separable state, where $\varphi$ can be any state. 
We know that the Generalized Robustness can be calculated from 
Eq.\ref{OEW}, using $\M{M} = \{W \in \M{W}\, |\, W \leq I\}$\cite{brandao05}, 
where  $I$ is the identity operator; 
in other words,
\begin{equation}
\M{R}_{g}(\rho) = max(0, -\min_{\{W \in \M{W}\, |\, W \leq I\}} Tr(W\rho)).
\label{rob.generalized}
\end{equation}

The construction of entanglement witnesses is a hard problem.
In an interesting  method proposed 
by Brand\~ao and Vianna   \cite{Reinaldo04b},  the optimization of 
entanglement  witnesses  is cast as a
{\em robust semidefinite program} (RSDP). 
Despite  RSDP is  computationally intractable, it is possible to 
perform a probabilistic relaxation turning it into   a semidefinite
program(SDP), which can be solved efficiently \cite{convexoptimization}.

\subsection{Quantum Correlations}
\label{quantum.correlations}

The space state for  indistinguishable fermions
is antisymmetric  under permutation of particles. 
In this case, it is convenient to use the 
{\em second quantization} formalism, in order to deal with the antisymmetric 
states in the {\em Fock space}. 
Accordingly  we introduce an algebra of operators which satisfy the 
following anti-commutation relations:
\begin{equation}
\{f^{\dagger}_i,f^{\dagger}_j\} = \{ f_i,f_j\} =  0, \qquad \{f_i,f^{\dagger}_j\} =\delta_{ij},
\label{fermionic.anti.comutation}
\end{equation}
where $f^{\dagger}_i$ and $f_i$ are the fermionic creation and 
annihilation operators, respectively, so that their application on 
the vacuum state ($\ket{0}$) creates/annihilates a fermion in state ``i''. 
The vacuum state is defined so that $f_i \ket{0} = 0$. 

An immediate and disturbing consequence of the antisymmetric structure
of the state space 
 can be seen even in the simplest example of a two-fermion system,  
which, if analysed in the usual way, will always  be considered   
entangled. 
We must therefore  rethink the way entanglement is calculated 
for systems of indistinguishable particles,
 as well as its  physical interpretation.

In the case which the identical particles are localized in 
\textit{distinct} laboratories and \textit{independently} prepared, 
it is natural to think  that the entanglement calculated in 
the usual way should not have any relevant physical meaning; or rather,
 ``no quantum prediction, referring 
to an atom located in our laboratory, is affected by the mere presence 
of similar atoms in remote parts of universe'' \cite{asherperezbook}.     

We are interested in the case of identical particles that are sufficiently 
close together such that the overlap between their wave functions is no
 longer negligible, and therefore they are indistinguishable.
Fermionic systems of this kind can be described using 
Slater determinants \cite{Kais}. 
Consider, for example, a  two-fermion state represented by a single
 Slater determinant,  namely,
\begin{equation}
\ket{\psi} = \frac{1}{\sqrt{2}}(\ket{\phi}\otimes \ket{\chi} - 
\ket{\chi}\otimes \ket{\phi}) = f^{\dagger}_{\phi} f^{\dagger}_{\chi}\,\ket{0},
\end{equation}
where $\ket{\phi}$ and $\ket{\chi}$ correspond to orthonormal wave functions
({\em spin-orbitals}). 
It is easy to see, in this simple case,
that the anti-symmetrization of coordinates introduces 
correlations between the fermions, namely, the well known
{\em exchange contributions} from
the Hartree-Fock theory. On the other hand, a single
Slater determinant is solution of a one-particle Schr\"odinger equation
and, therefore, can have no quantum correlation between the particles
\cite{Kais}.
Considering states described by more than one Slater determinant 
introduces  additional correlations beyond the exchange contribution.
We will then interpret such additional correlations as the analog of 
quantum entanglement in systems of distinguishable particles, calling 
them as {\em fermionic entanglement} \cite{Schliemann02}.

A measure of fermionic entanglement was proposed in \cite{Schliemann02} 
as the  analogous of  Wootters concurrence \cite{wootters98}. 
Notwithstanding,
 such measure, called Schliemann concurrence ($C_{S}$), is valid only for 
two-fermion states with a four-dimensional single-particle  Hilbert space 
($\M{A}(\Md{H}{4}\otimes\Md{H}{4})$), i.e.  the antisymmetric space of 
lowest dimension where can exist quantum correlated states.

In order to define the Schliemann concurrence, 
we have to introduce  some operators. 
Let  $\M{U}_{ph}$ be the operator of particle-hole transformation:
\begin{equation}
\M{U}_{ph} f_i^{\dagger} \M{U}_{ph}^{\dagger} = f_i, \quad \quad  
\M{U}_{ph}\ket{0} = \prod_{i=1}^{d} f_i^{\dagger} \ket{0},
\end{equation}
being $d$ the single-particle  Hilbert space dimension. 
Similarly, define $\M{K}$ as the anti-linear operator of 
complex-conjugation, satisfying the following relations:
\begin{equation}
\M{K} f_i^{\dagger} \M{K} = f_i^{\dagger}, \quad \quad 
\M{K} f_i \M{K} = f_i, \quad \quad \M{K} \ket{0} = \ket{0}.
\end{equation}
Thus, given the operator $\M{D} = \M{K}\M{U}_{ph}$, called 
operator of dualisation, and the dual states $\widetilde{\rho} = 
\M{D}\rho\M{D}^{-1}$, we have that the Schliemann concurrence for 
states $\rho \in \M{A}(\Md{H}{4}\otimes\Md{H}{4})$ is given by
\begin{equation}
C_{S}(\rho) = max(0 , \lambda_6 - \lambda_5 - \lambda_4 - \lambda_3 - \lambda_2 - \lambda_1),
\label{Schiemann.concurrence.fermions}
\end{equation}
where $\lambda_i's$ are, in descending order of magnitude, 
the square roots of the singular values of the matrix $R = \rho \widetilde{\rho}$.

\subsection{Quantum Discord}
\label{discord}

The total amount of  correlations (classical and quantum) 
 carried in a bipartite system 
is quantified by the mutual information:
\begin{equation} I(A:B)= S(A)+S(B)-S(A,B),\end{equation} where
 $S(X)$ is the Shannon (von Neumann) entropy  for the
classical (quantum) system. 
In the classical case, the mutual information has another interpretation:
 it measures
the decrease of ignorance about the subsystem $A$, when $B$ is known.
In this case,  an equivalent expression for the mutual information reads:
\begin{equation} J(A:B)=S(A)-S(A|B),\end{equation} where 
\begin{equation} S(A|B)= \sum_{b}p_{b} S(A|b)\end{equation} is the conditional entropy on $A$ given $B$.
 In quantum systems this equivalence is not
 always true. The disagreement between these expressions is
 called \textit{quantum discord}, and can be used to quantify
 quantum correlations \cite{oz2001}:
\begin{equation}
D(A:B)=I(A:B)-J(A:B).
\end{equation}
However, there is an ambiguity in $J(A:B)$ due to the freedom of 
choice for the measurement operators, where for each set of measurements 
$\{\Pi_{i} \}_B$ on B, the conditional entropy value may be different 
and consequently also the mutual information. 
The quantum discord is thus redefined by minimizing its value over all
 measurement operators. The mutual information is thus properly defined
by:
\begin{equation}
J(A:B) = S(A)-\max\limits_{\{\Pi_{i} \}_B} S(A|B),
\end{equation}
and it is known to quantify classical correlations \cite{hv2001}.
 Thus, the quantum discord measures the difference between the total and 
classical correlations on the system, therefore being  a quantifier of 
 quantum correlations. Note however that quantum discord is not
a measure of entanglement, and quantum separable states usually have 
non-zero discord. The quantum states with null discord are those
which are a mere  encoding of classical statistical distributions,
and can be written in the form:
\begin{equation}\label{discord.distinguishable}
\xi = \sum_{ij}\lambda_{ij} \ketbra{e_i}{e_i}\otimes\ketbra{f_j}{f_j}.
\end{equation}
where $\ket{e_i}$ and $\ket{f_j}$ are orthonormal basis.

Given the set ($\Omega$) of states with zero-discord, formed by the 
states above, it is clear that the minimal distance between a state 
$\rho$ and this set can be used to quantify the quantum discord 
\cite{bm2011}:
\begin{equation}
\M{D}(\rho)=\min_{\xi\in\Omega}|\xi-\rho|_{p},
\label{geom.discord.distinguishable}
\end{equation}
where $|A|_{p}$ is the Schatten $p-norm$. A usual measure of quantum discord is based on the 
$2$-norm, or Hilbert-Schmidt norm, proposed by Daki\'c \textit{et al.}
 \cite{dvb2010}.

\section{Quantifying Quantum Correlations using Witness Operators}
\label{quantifying.quantum.correl}

In this section we will present our method for quantifying 
fermionic entanglement using witness operators. 
After defining the set of separable states pertinent to our case,
i.e. the  fermionic states without entanglement, 
we will introduce a 
method for determining  OEWs. 
In particular, we will see which constraints to impose on the 
set $\M{M}$ of witnesses (Eq.\ref{OEW}) in order to obtain a 
quantifier analogous to the Generalized Robustness (Eq.\ref{rob.generalized}).
We also will define
a notion of quantum discord for fermions and confront it to entanglement.

We know that states without entanglement are those that 
can be described by a single Slater determinant, or a convex 
mixture of them.
Consider  $\M{F}_n^d$ as the Fock space of $n$ 
indistinguishable fermions sharing a $d$-dimensional single-particle 
Hilbert space.  We have then the following definition of ``separable'' states:\\

\textit{\textbf{State with no fermionic entanglement (separable):}} 
A state $\sigma \in \M{B}(\M{F}_n^d)$ has no fermionic entanglement
if  it can be decomposed as
\begin{equation}
\sigma = \sum\limits_i p_i \,\,a^{i^{\displaystyle\dagger}}_{1} 
\cdots \, a^{i^{\displaystyle\dagger}}_{n} \ket{0}\bra{0}\, a^{i}_{n} 
\cdots\, a^{i}_{1}, \quad\sum\limits_i p_i = 1,
\label{separable.state.fock}
\end{equation}
where $a^{i^{\displaystyle\dagger}}_{k} = \sum\limits_{l=1}^{d} c_{l}^{ik} 
\, f^{\displaystyle\dagger}_{l}$, and $\{ f^{\displaystyle\dagger}_{l}\}$ 
is an orthonormal basis of fermionic creation operators for the space 
of a single fermion ($\M{F}_1^d$).
Of course, the states defined by Eq.\ref{separable.state.fock} are not
separable in the usual mathematical sense, meaning that they  are
 product states or convex mixtures of it. But we will insist in 
referring to them as {\em separable}, for they are just 
anti-symmetrization of the usual distinguishable separable states.
Entanglement, in the case of distinguishable particles, is defined
in opposition to separability, i.e., an entangled state is that one
which is not separable. We want to keep this notion. 

It is interesting to note that, as in the case of 
distinguishable particles, the set of separable states is 
invariant under {\em local operations}, taking now into account 
that the local 
operations must be symmetric, due to  the indistinguishability of 
the particles.  
Let $\Phi$ be a {\em local symmetric operation} (LSO), i.e., an operation 
that respects the Pauli exclusion principle and does not involve any
interaction between particles. An LSO can be written as:
\begin{equation}
\Phi(\rho) = \sum\limits_{i} (M_i \otimes M_i \otimes \cdots \otimes M_i)\, 
\rho \,(M_i^{\dagger} \otimes M_i^{\dagger} \otimes \cdots \otimes M_i^{\dagger})
\end{equation}
where $M_i$ is a  linear operator acting on the Hilbert space of a single 
particle. Given a fermionic separable pure state
 (i.e. a single Slater determinant)
 $\ket{\psi_{sep}} = \M{A} (\ket{e_1} \otimes \cdots \otimes \ket{e_n})$, where $\M{A}$ is the anti-symmetrization operator, $\{\ket{e_i}\}$ is an orthonormal basis, and noting that $[\Phi,\M{A}] = 0$, we see that
\begin{eqnarray}
\nonumber (M_i \otimes \cdots \otimes M_i)\, 
\ket{\psi_{Sep}} &=& (M_i \otimes \cdots \otimes M_i) 
\, \M{A} \ket{e_1\cdots e_n}\\ 
\nonumber &=& \M{A} \, (M_i \otimes \cdots \otimes M_i) 
\ket{e_1\cdots e_n}\\ 
&=& \M{A} \, \ket{(M_{i}e_1) \cdots (M_{i}e_n)} \nonumber\\
 &=& \M{A} \, \ket{e_1'\cdots e_n'}\label{op.LOCS.sep.pure}\\
\nonumber &=& \ket{\psi_{Sep}'}
\end{eqnarray}
and such result clearly extends to mixed states. Summarizing, given 
a separable state $\sigma \in \M{S}$, we have that 
$\Phi_{LSO}(\sigma) \in \M{S}$, indicating that 
in order to have quantum entanglement, the particles must interact by
means of some global operation.

Now we adapt  Brand\~ao and Vianna's \cite{Reinaldo04b} technique in 
order to obtain a new algorithm to determine OEWs for indistinguishable 
fermions in the Fock space. The new method can be enunciated as follows.

\textit{\textbf{Determination of OEW using RSDP:}} A fermionic state 
$\rho \in \M{B}(\M{F}_n^d)$  is entangled
if and only if  the optimal value of the following RSDP is negative:
\begin{center}
{\em minimize $Tr(W\rho)$ subject to}
\begin{equation}
\left\{
\begin{array}{c}
\sum\limits_{i_{n-1}=1}^{d} \cdots \sum\limits_{i_1=1}^{d} 
\sum\limits_{j_1=1}^{d} \cdots \sum\limits_{j_{n-1}=1}^{d} 
(c_{i_{n-1}}^{n-1^*} \cdots c_{i_1}^{1^*} \, c_{j_1}^1 \cdots c_{j_{n-1}}^{n-1} \times \,\\
  W_{i_{n-1} \cdots i_1 \, j_1 \cdots j_{n-1}}) \geq 0,  \\
\forall c_{i}^k \in \M{C}, \,\, 1\leq k \leq (n-1), \,\, 1\leq i \leq d, \\
\M{A}W\M{A}^{\dagger}=W, \\
W\leq \M{A}, 
\end{array} 
\right.
\label{RSDP}
\end{equation}
\end{center}
where $d$ is the dimension of the  single particle Hilbert space, 
$\{ f^{\dagger}_{l}\}$ is an orthonormal basis 
of fermionic creation operators, $\M{A}$ is the anti-symmetrization operator,
and $W_{i_{n-1} 
\cdots i_1 \, j_1 \cdots j_{n-1}} = f_{i_{n-1}} \cdots f_{i_1}\, 
W \,f^{\dagger}_{j_1} \cdots f^{\dagger}_{j_{n-1}} \in 
\M{B}(\M{F}_1^d)$ 
is an operator acting on the space of one fermion.
The notation $W\leq \M{A}$ means 
that $(\M{A} - W) \geq 0$ is a positive semidefinite operator. 
If $\rho$ is entangled, the operator $W$ that minimizes 
the problem corresponds to the OEW of $\rho$. \\

\textit{\textbf{Proof:}} It is known that a state is entangled if and only if 
there exists a witness operator $W$ such that $Tr(W\rho)<0$  and 
$Tr(W\sigma)\geq 0$ for every separable state $\sigma$. 
Consider a general separable state as given by Eq.\ref{separable.state.fock}. 
The semi-positivity condition  $Tr(W\sigma)\geq 0$  is equivalent to:
\begin{equation}
\label{aWa}
\bra{0} a_{n} a_{n-1} \cdots a_{1}\, W \, a^{\dagger}_{1} 
\cdots a^{\dagger}_{n-1} a^{\dagger}_{n} \ket{0} \geq 0,
\end{equation}    
for all $a^{\dagger}_{k} \in \M{B}(\M{F}_{1}^{d}).$
Note however that to satisfy Eq.\ref{aWa}, it is sufficient that 
the operator $a_{n-1} \cdots a_{1}\, W \, a^{\dagger}_{1} 
 \cdots a^{\dagger}_{n-1}$ be positive semidefinite.
Thus follows directly that the operator $W$ satisfying 
the problem in Eq.\ref{RSDP} corresponds to an optimal entanglement 
witness.  \\

The  RSDP given above is solved by means of probabilistic relaxations
it terms of SDPs,
as done in \cite{Reinaldo04b}, where the set of infinite constraints
is exchanged by a finite sample.
Thus the  witness operator obtained is such that satisfy  most of the
constraints in Eq.\ref{RSDP}.  
The small probability ($\epsilon$) that  a constraint be violated
(i.e. $Tr(W\sigma)<0$) diminishes  as the size of the sample of constraints
increases.

The constraint $\M{A}W\M{A}^{\dagger}=W$ restricts the operator to 
the space of antisymmetric entanglement witnesses 
($\M{W}(\M{F}_n^d) = A\M{W}A^{\dagger}$). The other constraint,
$W\leq \M{A}$, follows directly from the anti-symmetrization of
Eq.\ref{rob.generalized}, and implies that the OEW corresponds to
the anti-symmetrized version of the Generalized Robustness, namely,
\begin{equation}
\M{R}_{g}^{\M{F}}(\rho) = max(0, -\min_{\M{M} = \{W \in 
\M{W}(\M{F}_n^d)\, |\, W \leq \M{A}\}} Tr(W\rho)).
\label{rob.generalized.fermionic}
\end{equation} 
$\M{R}_{g}^{\M{F}}(\rho)$  measures the minimum required mixture
 with a fermionic state such that all the entanglement of
$\rho$ is washed out. In other words,
the Generalized Robustness is the minimum value of $s$ such that
\begin{equation}
\sigma=\frac{\rho + s\varphi_f}{1+s}
\label{ferm.rob.generalized.definition}
\end{equation} 
be a separable state (Eq.\ref{separable.state.fock}), where $\varphi_f$ 
can be any fermionic state.

\begin{figure}
\includegraphics[scale=0.44]{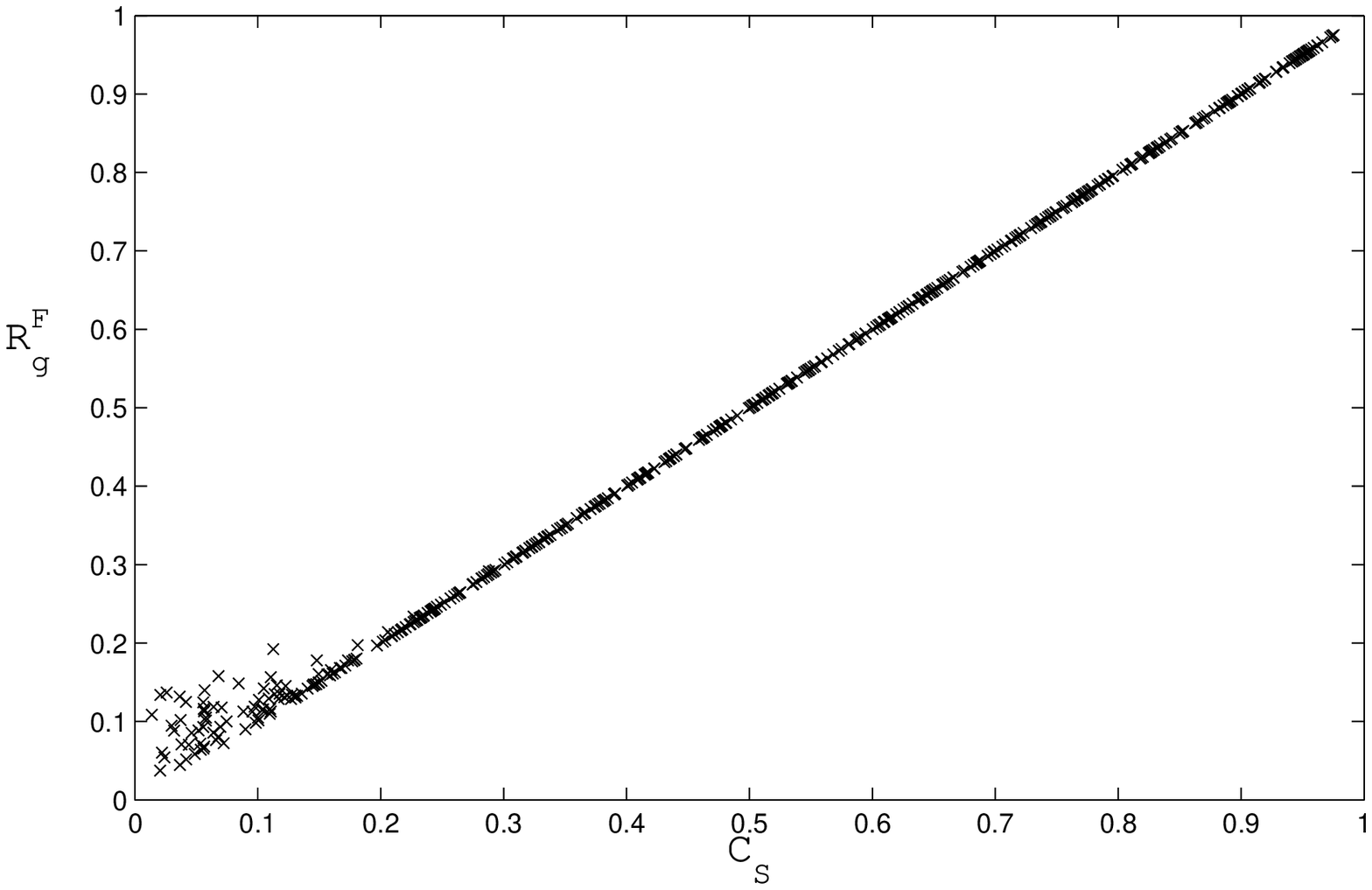}
\includegraphics[scale=0.44]{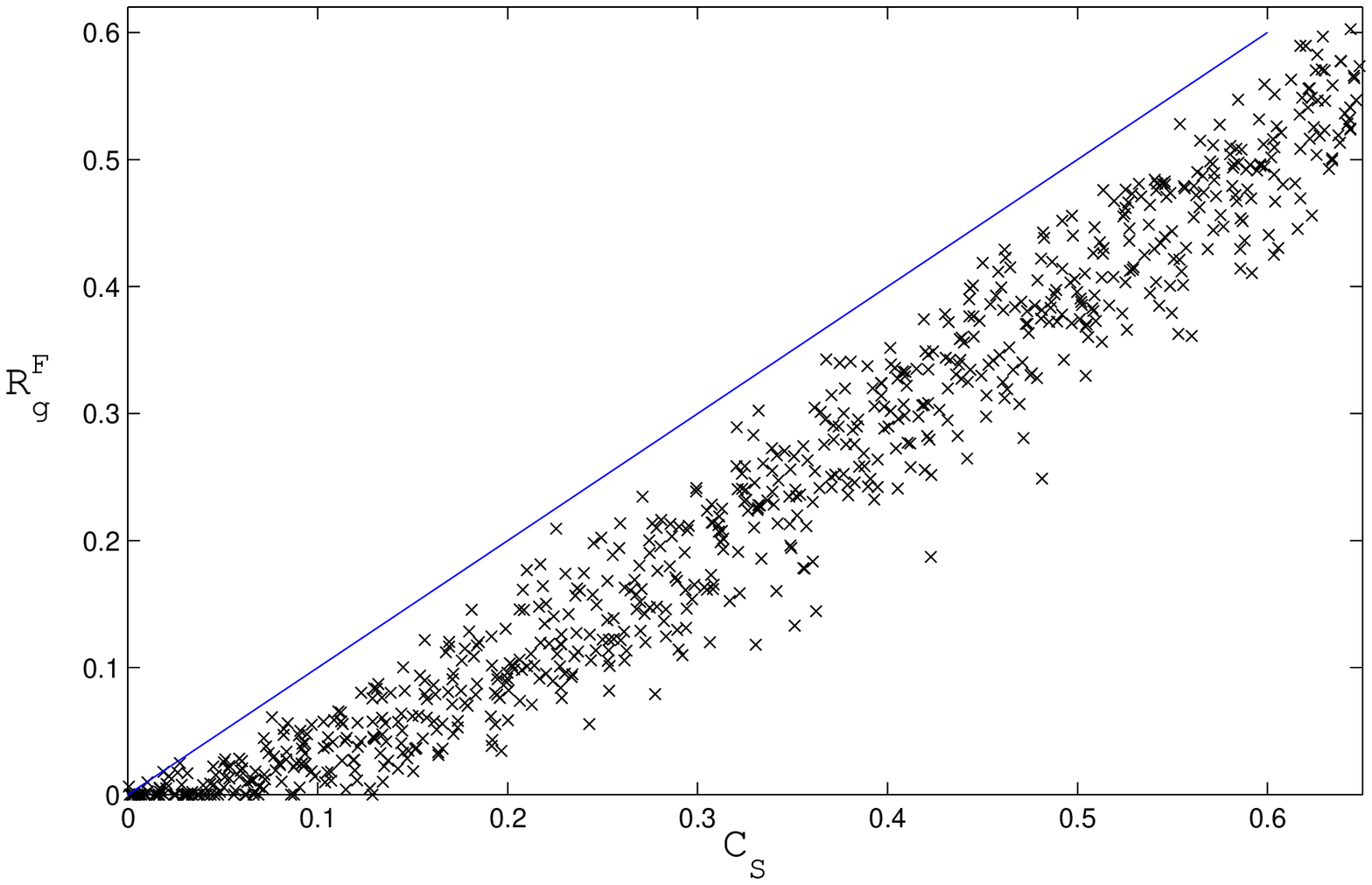}
\caption{ Fermionic 
Generalized Robustness $\M{R}_g^\M{F}$ versus
Schliemann Concurrence $C_S$, for random  fermionic states uniformly
distributed according to the Haar measure.
(TOP) The two entanglement measures are equal for pure states.
 The small dispersion seen in the
top panel is due to numerical imprecision in the calculation of
 $\M{R}_g^\M{F}$ in the region of very low entanglement.
(BOTTOM) In the case of mixed states, the  $C_S$ is an upper bound
to $\M{R}_g^\M{F}$.  The continuous line in the bottom panel corresponds
to the straight line  $C_S= \M{R}_g^\M{F}$, and is just a guide to the
eye.} 
\label{concurrenceXrobustness}
\end{figure}

\begin{figure}
\includegraphics[scale=0.44]{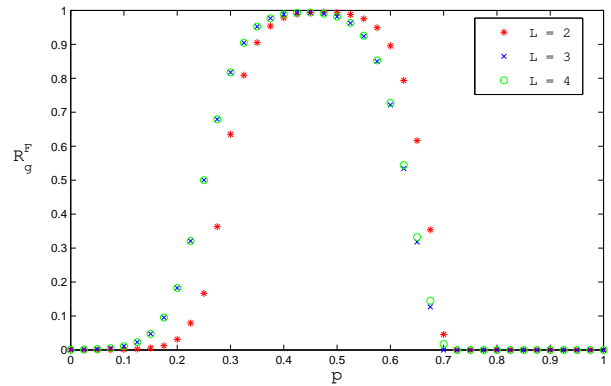}
\caption{Fermionic Generalized Robustness 
$\M{R}_ g^\M{F}$ for the families of two-fermion 
states defined in Eq.\ref{family.states}, on the space 
$\M{A}(\Md{H}{2L}\otimes\Md{H}{2L})$.} 
\label{familystatesN2}
\end{figure}

 \textit{\textbf{Fermionic Geometric Discord:}} 
Abiding by  the notion that  mere anti-symmetrization 
does not generate any 
kind of quantum correlation, 
we are led to the following definition of fermionic states without
 quantum discord:
\begin{equation}
\xi_{A} = \M{A} \,\xi \,\M{A}^{\dagger},
\label{fermionic.states.no.discord}
\end{equation}   
where $\M{A}$ is the anti-symmetrization operator, and $\xi$ are the
states in Eq.\ref{discord.distinguishable}, 
 which encode classical probability distributions.  
Although these states do not have any kind of quantum correlations, they cannot be 
treated like  classical probability distributions, since they respect 
quantum rules: like the Pauli exclusion principle.

Our proposed measure for the quantum discord in fermionic states will be a 
geometric measure like Eq.\ref{geom.discord.distinguishable}. 
Given a fermionic state $\rho\in \M{A}(\Md{H}{d}\otimes\Md{H}{d})$, the Fermionic
 Geometric Discord is given by,
\begin{equation}
\M{D}_f(\rho)= \min_{\xi\in\Omega_{A}}|\xi-\rho|_{1},
\label{geom.fermionic.disc}
\end{equation}
where $\Omega_{A}=\M{A}\,\Omega \M{A}^{\dagger}$ 
is the set of zero-discord anti-symmetric states (Eq.\ref{fermionic.states.no.discord}).

\section{Numerical Results}
\label{numerical.results}

In this section we will illustrate our method.
We start by investigating {\em bipartite entanglement}
 in the space  $\M{A}(\Md{H}{4}\otimes\Md{H}{4})$,
which has the  smallest dimension allowing
for quantum correlations in fermionic systems.
In this case, we can compare  the Fermionic Generalized Robustness
($\M{R}_g^\M{F}$ - Eq.\ref{rob.generalized.fermionic}) with 
the Schliemann concurrence ($C_S$ - Eq.\ref{Schiemann.concurrence.fermions}).
 Then we investigate
the bipartite entanglement in a one-parameter family of states in the space
$\M{A}(\Md{H}{2L}\otimes\Md{H}{2L})$, with $L$ going from 2 to 4.
We also compare our  Fermionic Geometric Discord 
($\M{D}_f$ - Eq.\ref{geom.fermionic.disc}) with
 the Fermionic Generalized Robustness
for another one-parameter family of states in the
 space $\M{A}(\Md{H}{4}\otimes\Md{H}{4})$.
 We finish with calculations of {\em multipartite entanglement} in the
Extended Hubbard Model (EHM), where $\M{R}_g^\M{F}$ can characterize 
{\em quantum phase transitions}.

In Fig.1, we plot  the Fermionic Generalized Robustness against 
the  Schliemann concurrence  for states generated randomly
in the space  $\M{A}(\Md{H}{4}\otimes\Md{H}{4})$.
It is remarkable that  $C_S$ and $\M{R}_g^\M{F}$ are one and
the same for pure states, and  $C_S$ is an upper bound for $\M{R}_g^\M{F}$
in the case of mixed states. Recall that, in the case of  two distinguishable 
qubits,  the Wootters Concurrence \cite{wootters98}  and the 
Generalized Robustness \cite{vidal99} also keep this same relation.

Now we consider the following one-parameter family of states, 
in the space $\M{A}(\Md{H}{2L}\otimes\Md{H}{2L})$:
\begin{equation}
\rho = f(0)\sigma  + f(1/2)\rho_{max}  + f(1)\M{A}, \quad f(x) = 
Ae^{\displaystyle\frac{-(p - x)^2}{\Delta^2}}, 
\label{family.states}
\end{equation}
where $\sigma = f^{\dagger}_{i} f^{\dagger}_{j} \ket{0}$ ($i\neq j$) 
is a pure state with just a single Slater determinant; $\rho_{max}$ 
is the maximally entangled pure state of singlet type, i.e.
the one with spin quantum numbers
$S=S_z=0$; and $\M{A}$ is the anti-symmetrizer, which corresponds 
to the identity operator  in the antisymmetric space. 
$p$ controls the entanglement of the state, $A$ is  chosen such that the
state is normalized, and $\Delta = 0.1826$.
  In Fig.2, the Fermionic Generalized Robustness is 
plotted against $p$, for spaces of different dimensions. We see that
 $\M{R}_g^\M{F}$ behaves correctly, showing that the mixed state has 
much entanglement when the contribution of the singlet is large, 
and it has low, or none, entanglement when the contribution of 
either $\sigma$ or $\M{A}$ is large.

\begin{figure}
\includegraphics[scale=0.36]{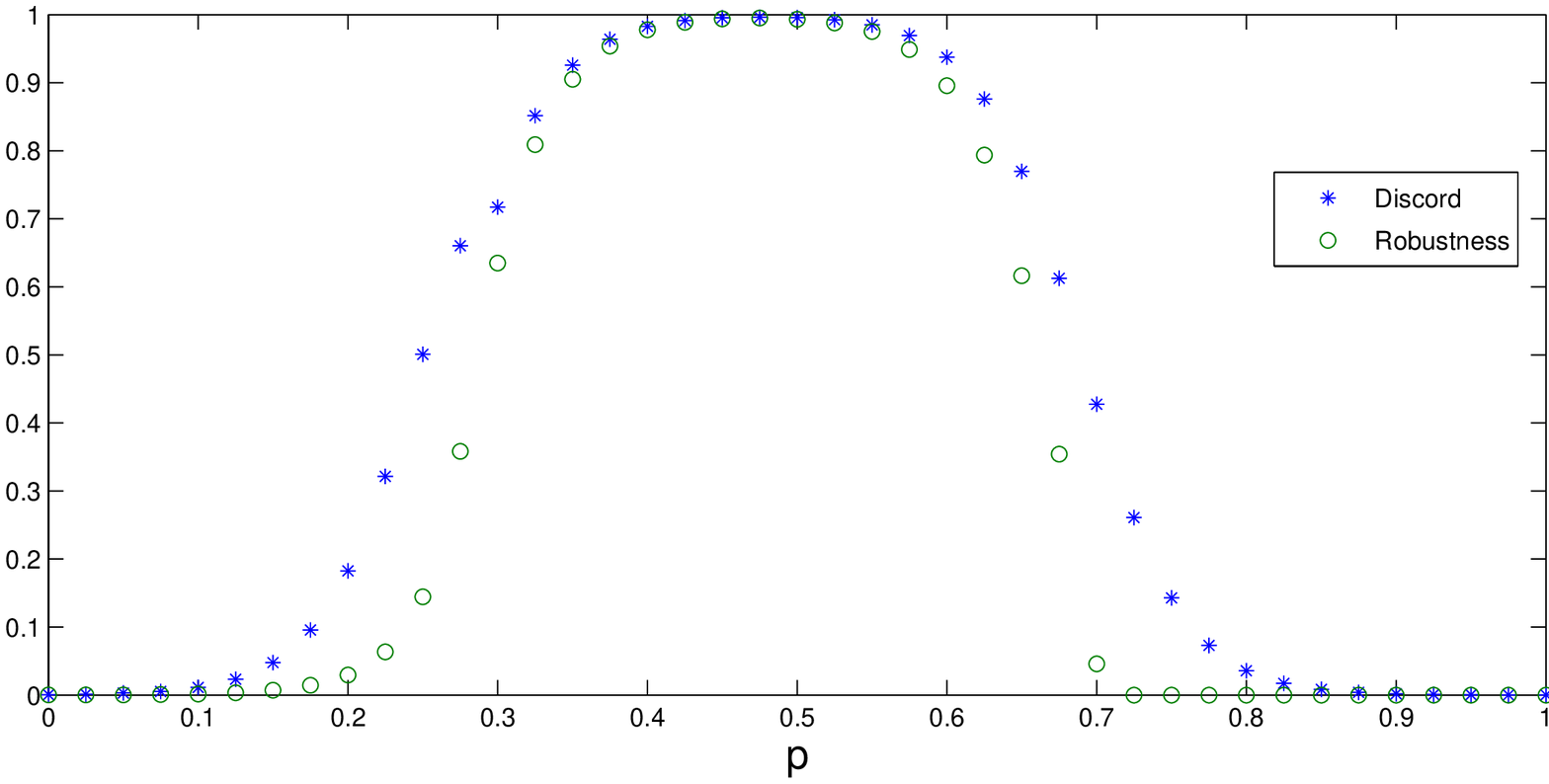} 
\includegraphics[scale=0.40]{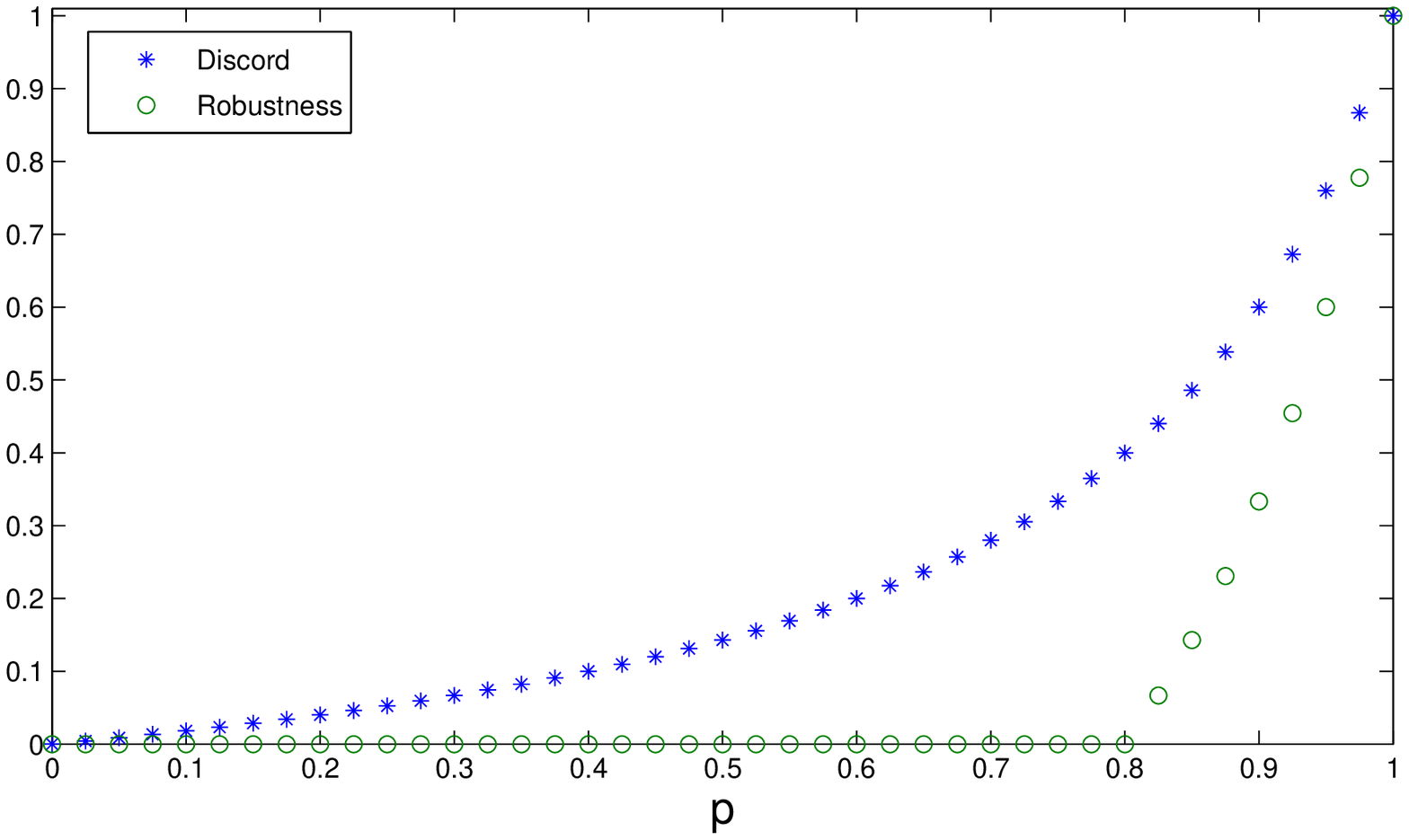}
\caption{Fermionic Generalized Robustness 
$\M{R}_ g^\M{F}$, and Fermionic Geometric Discord $\M{D}_f$ for the families of two-fermion 
states defined in Eq.\ref{family.states} (top) and Eq.\ref{family.state.two} (bottom), on the space 
$\M{A}(\Md{H}{4}\otimes\Md{H}{4})$.} 
\label{fig.robustness.discord}
\end{figure}

\begin{figure}
\includegraphics[scale=0.43]{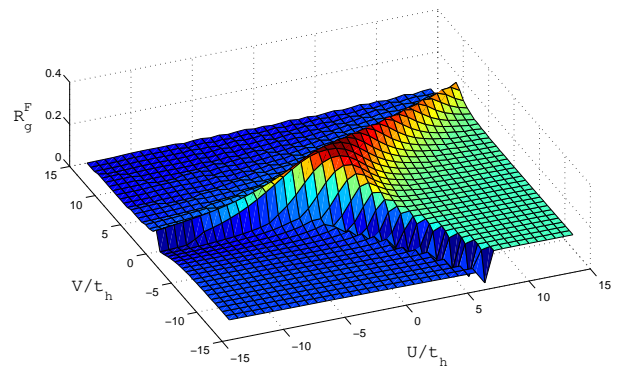}
\caption{Five-partite Fermionic Generalized Robustness
in the Extended Hubbard Model (Eq.19). The multipartite entanglement
works as an order parameter and characterizes quantum 
phase transitions, in this case, with three distinct phases.}
\end{figure}

Consider another one-parameter family of states, in the space $\M{A}(\Md{H}{4}\otimes\Md{H}{4})$:
\begin{equation}
\rho = (1-p)\M{A} + p\rho_{max}, \quad 0\leq p\leq 1,
\label{family.state.two}
\end{equation}
In Fig.3, we confront the Fermionic Generalized Robustness and the Fermionic
 Geometric Discord, for the two families of states (Eq.26, Eq.27) defined above.
In the two cases, we see that the discord is always an upper bound for the
entanglement, and states without entanglement can have a non-null discord, 
which is particularly dramatic in the second family.
In the first family of states, the functional forms for  the discord and
entanglement are very similar, whereas they are very different in the second
family. Note, in the second family,  the abrupt vanishing of entanglement
(a discontinuous derivative for $p\cong 0.8$),
while the discord shows an asymptotic behavior.

To conclude, we illustrate the calculation of multipartite entanglement
of fermions interacting according to  the Extended Hubbard Model \cite{EHM},
defined by the Hamiltonian:
\begin{equation}
\begin{array}{r}
H_{EHM}=-t_h\sum_{j=1,\,\sigma = \uparrow, \downarrow}^{L}
(f^{\dagger}_{j,\sigma}f_{j+1,\sigma} + f^{\dagger}_{j+1,\sigma}f_{j,\sigma}) 
+ \\
 \mbox{U}\sum_{j=1}^{L}n_{j\uparrow}n_{j\downarrow} + \mbox{V} \sum_{j=1}^{L}n_{j}n_{j+1},
\end{array}
\end{equation}
where $U$ and $V$ define the on-site and nearest-neighbor Coulomb interactions,
$t_h$ controls hopping between sites,  $L$ is the number of sites, and
$n_{j\sigma}$ is  the particle number operator on site $j$ with spin $\sigma$.
Fig.4 shows the five-partite Fermionic Generalized Robustness of the 
ground state as a function
of $U/t_h$ and $V/t_h$, for the case of five fermions and five sites.
The ground state is obtained by numerical diagonalization of the Hamiltonian.
 One can see that the Fermionic Generalized Robustness characterizes
 three distinct regions, 
corresponding exactly to the three distinct phases provided
 by the known phase diagram of the model, namely charge-density wave (up),
 spin-density wave (right) and phase separation (bottom) \cite{EHM}. 
It is interesting to note that five fermions in five sites, with periodic
boundary conditions, is the smallest size of the system which presents 
such phase transitions. The occurrence of these phase transitions is
dependent on long range interactions. Surprisingly, the second-neighbor 
interactions present in our  five fermions model is already 
{\em long range} enough for the onset of  phase transitions.
We performed a calculation with four fermions in four sites, and the
resulting figure is a uninteresting flat  surface for the entanglement.

\section{Conclusion}
\label{conclusion}

In summary, we  presented  a method to quantify 
quantum correlations in systems of fermionic indistinguishable particles. 
The method is based on the use of optimal entanglement witnesses,
which can be calculated with arbitrary precision  by means of SDPs.
In particular, we obtained the  Generalized Robustness for 
fermionic systems ($\M{R}_{g}^\M{F}$), and numerically showed its
relation to the  Schliemann concurrence.
We also introduced the Fermionic Geometric
 Discord ($\M{D}_f$),
and observed that it is an upper bound for the fermionic entanglement. 
However, the physical meaning of quantum discord for fermionic systems
needs to be better understood. It is comforting to know that the
quantum discord for a single Slater determinant, or for the fermionic
maximally mixed state, is null, but the nature  of quantum correlations
 conveyed by a convex mixture of Slater determinants is still obscure to
 us,
and will be investigated in future works. 
 Finally, we used 
the five-partite Fermionic Generalized
Robustness to characterize quantum phase transitions in the Extended
Hubbard Model, showing the utility of entanglement as a quantum order
parameter.

{\em Acknowledgments} - Financial support by the
Brazilian agencies  FAPEMIG, CNPq, and  INCT-IQ (National 
Institute of Science and Technology for Quantum Information).


\end{document}